\documentclass[aps,prl,twocolumn,superscriptaddress,showpacs,preprintnumbers,amsmath,amssymb]{revtex4}
\usepackage{graphicx}
\usepackage{dcolumn}
\usepackage{bm}
\usepackage{amsmath}
\usepackage{amssymb}
\usepackage{color}

\begin{document}

\preprint{\vbox{ 
								 \hbox{BELLE-CONF-0822}
                              }}

\title{\quad\\
[0.5cm] Measurement of the Differential Branching Fraction 
and Forward-Backward Asymmetry for $B \to K^{(*)} \ell^+ \ell^-$}

\affiliation{Budker Institute of Nuclear Physics, Novosibirsk}
\affiliation{Chiba University, Chiba}
\affiliation{University of Cincinnati, Cincinnati, Ohio 45221}
\affiliation{Department of Physics, Fu Jen Catholic University, Taipei}
\affiliation{Justus-Liebig-Universit\"at Gie\ss{}en, Gie\ss{}en}
\affiliation{The Graduate University for Advanced Studies, Hayama}
\affiliation{Gyeongsang National University, Chinju}
\affiliation{Hanyang University, Seoul}
\affiliation{University of Hawaii, Honolulu, Hawaii 96822}
\affiliation{High Energy Accelerator Research Organization (KEK), Tsukuba}
\affiliation{Hiroshima Institute of Technology, Hiroshima}
\affiliation{University of Illinois at Urbana-Champaign, Urbana, Illinois 61801}
\affiliation{Institute of High Energy Physics, Chinese Academy of Sciences, Beijing}
\affiliation{Institute of High Energy Physics, Vienna}
\affiliation{Institute of High Energy Physics, Protvino}
\affiliation{Institute for Theoretical and Experimental Physics, Moscow}
\affiliation{J. Stefan Institute, Ljubljana}
\affiliation{Kanagawa University, Yokohama}
\affiliation{Korea University, Seoul}
\affiliation{Kyoto University, Kyoto}
\affiliation{Kyungpook National University, Taegu}
\affiliation{\'Ecole Polytechnique F\'ed\'erale de Lausanne (EPFL), Lausanne}
\affiliation{Faculty of Mathematics and Physics, University of Ljubljana, Ljubljana}
\affiliation{University of Maribor, Maribor}
\affiliation{University of Melbourne, School of Physics, Victoria 3010}
\affiliation{Nagoya University, Nagoya}
\affiliation{Nara Women's University, Nara}
\affiliation{National Central University, Chung-li}
\affiliation{National United University, Miao Li}
\affiliation{Department of Physics, National Taiwan University, Taipei}
\affiliation{H. Niewodniczanski Institute of Nuclear Physics, Krakow}
\affiliation{Nippon Dental University, Niigata}
\affiliation{Niigata University, Niigata}
\affiliation{University of Nova Gorica, Nova Gorica}
\affiliation{Osaka City University, Osaka}
\affiliation{Osaka University, Osaka}
\affiliation{Panjab University, Chandigarh}
\affiliation{Peking University, Beijing}
\affiliation{Princeton University, Princeton, New Jersey 08544}
\affiliation{RIKEN BNL Research Center, Upton, New York 11973}
\affiliation{Saga University, Saga}
\affiliation{University of Science and Technology of China, Hefei}
\affiliation{Seoul National University, Seoul}
\affiliation{Shinshu University, Nagano}
\affiliation{Sungkyunkwan University, Suwon}
\affiliation{University of Sydney, Sydney, New South Wales}
\affiliation{Tata Institute of Fundamental Research, Mumbai}
\affiliation{Toho University, Funabashi}
\affiliation{Tohoku Gakuin University, Tagajo}
\affiliation{Tohoku University, Sendai}
\affiliation{Department of Physics, University of Tokyo, Tokyo}
\affiliation{Tokyo Institute of Technology, Tokyo}
\affiliation{Tokyo Metropolitan University, Tokyo}
\affiliation{Tokyo University of Agriculture and Technology, Tokyo}
\affiliation{Toyama National College of Maritime Technology, Toyama}
\affiliation{Virginia Polytechnic Institute and State University, Blacksburg, Virginia 24061}
\affiliation{Yonsei University, Seoul}
  \author{I.~Adachi}\affiliation{High Energy Accelerator Research Organization (KEK), Tsukuba} 
  \author{H.~Aihara}\affiliation{Department of Physics, University of Tokyo, Tokyo} 
  \author{D.~Anipko}\affiliation{Budker Institute of Nuclear Physics, Novosibirsk} 
  \author{K.~Arinstein}\affiliation{Budker Institute of Nuclear Physics, Novosibirsk} 
  \author{T.~Aso}\affiliation{Toyama National College of Maritime Technology, Toyama} 
  \author{V.~Aulchenko}\affiliation{Budker Institute of Nuclear Physics, Novosibirsk} 
  \author{T.~Aushev}\affiliation{\'Ecole Polytechnique F\'ed\'erale de Lausanne (EPFL), Lausanne}\affiliation{Institute for Theoretical and Experimental Physics, Moscow} 
  \author{T.~Aziz}\affiliation{Tata Institute of Fundamental Research, Mumbai} 
  \author{S.~Bahinipati}\affiliation{University of Cincinnati, Cincinnati, Ohio 45221} 
  \author{A.~M.~Bakich}\affiliation{University of Sydney, Sydney, New South Wales} 
% ...
\collaboration{The Belle Collaboration}

\begin{abstract}
We study $B \to K^{(*)} \ell^+ \ell^-$ decays based on a large data sample 
of 657 million $B\overline{B}$ pairs collected with the Belle detector at the KEKB $e^+e^-$ collider. 
The differential branching fraction, the isospin asymmetry, 
the $K^{*}$ polarization, and the forward-backward asymmetry ($A_{FB}$) 
as functions of $q^2$ are reported. 
The fitted $A_{FB}$ spectrum tends to be shifted toward the positive side from the SM expectation. 
The measured branching fractions and lepton flavor ratios (electron/muon) are 
$\mathcal{B}$($B \to K^* \ell^+ \ell^-$) = (10.8$^{+1.1}_{-1.0}\pm0.9) \times 10^{-7}$, 
$\mathcal{B}$($B \to K \ell^+ \ell^-$) = (4.8$^{+0.5}_{-0.4}\pm0.3) \times 10^{-7}$, 
$R_{K^*}$ = 1.21$\pm0.25\pm0.08$, and $R_{K}$ = 0.97$\pm0.18\pm0.06$, respectively.

\end{abstract}

\collaboration{Belle Collaboration} \noaffiliation

\maketitle
    
\tighten    
The $b \to s \ell^+\ell^-$ transition is a flavor-changing neutral current (FCNC) process, 
which, in the Standard Model (SM), proceeds via either a 
$Z$/$\gamma$ penguin or a box diagram at lowest order. 
The effective Wilson coefficients $C_7$, $C_9$, and $C_{10}$ 
describe the amplitudes from the electromagnetic penguin, 
the vector electroweak, and the axial-vector electroweak contributions, respectively.
These amplitudes may interfere with the contributions from non-SM particles~\cite{susy}. 
Therefore, the branching fraction and the lepton forward-backward asymmetry ($A_{FB}$)
as functions of dilepton invariant mass 
in $b \to s \ell^+\ell^-$ provide information on the 
coefficients associated with certain theoretical models~\cite{kll_th}, 
and are also sensitive probes for the presence of new physics.

In this paper, we present measurements of 
the differential branching fractions and the isospin asymmetries 
as functions of $q^2=M_{\ell\ell}^{2} c^2$ for 
$B \to K^{*} \ell^+\ell^-$ and $B \to K \ell^+\ell^-$ decays.
The $K^{*}$ polarization and $A_{FB}$ for $B \to K^{*} \ell^+\ell^-$ decays 
as functions of $q^2$ are presented as well.
A data sample of 657 million $B\overline{B}$ pairs 
collected with the Belle detector at the KEKB $e^+e^-$ collider is examined. 
Charge-conjugate decays are implied throughout the paper. 
Equal production of $B^0\overline{B}{}^0$ and $B^+B^-$ pairs are assumed 
throughout this paper.


The Belle detector is a large-solid-angle magnetic spectrometer located at
the KEKB collider~\cite{KEKB}, and consists of a silicon vertex detector
(SVD), a 50-layer central drift chamber (CDC), an array of 
aerogel threshold Cherenkov counters (ACC), a barrel-like arrangement of
time-of-flight scintillation counters (TOF), and an electromagnetic
calorimeter (ECL) comprised of CsI(Tl) crystals located inside a
superconducting solenoid that provides a 1.5~T magnetic field. An iron
flux-return located outside the coil is instrumented to detect $K_L^0$
mesons and to identify muons (KLM). The detector is described in detail
elsewhere~\cite{belle_detector}.

We reconstruct $B \to K^{(*)} \ell^+\ell^-$ signal events in 10 final states: 
$K^{+} \pi^{-}$, $K_{S}^{0} \pi^{+}$, $K^{+} \pi^{0}$, $K^+$, and $K_{S}^{0}$ for $K^{(*)}$, 
and combine with either electron or muon pairs.
All charged tracks other than the $K_{S}^{0}\to\pi^+\pi^-$ daughters 
are required to have a maximum distance to the interaction point (IP) 
of 5 cm along the beam direction ($z$) and
0.5 cm in the transverse plane ($r$--$\phi$).
A track is identified as a $K^{+}$ ($\pi^{+}$) if 
the kaon likelihood ratio is greater (less) than 0.6 (0.4);
the kaon likelihood ratio is defined by 
$\mathcal{R}_{K}\equiv \mathcal{L}_{K}/(\mathcal{L}_{K}+\mathcal{L}_{\pi})$,
where $\mathcal{L}_{K}$ ($\mathcal{L}_{\pi }$) denotes a likelihood
that combines measurements from the ACC, the TOF, 
and $dE/dx$ from the CDC for the $K^{+}$ ($\pi^{+}$) tracks.
This selection is about 85\% (89\%) efficient for kaons (pions) while 
removing about 97\% (91\%) of pions (kaons).
In addition to the information included in the kaon likelihood ratio, 
muon (electron) candidates are
required to be associated with KLM detector hits (ECL calorimeter showers).
We define the likelihood ratio $\mathcal{R}_{x}$ ($x$ denotes $\mu$ or $e$) as
$\mathcal{R}_{x}\equiv \mathcal{L}_{x}/({\mathcal{L}_{x}}+{\mathcal{L}_{not-x}})$,
where $\mathcal{L}_{x}$ and $\mathcal{L}_{not-x}$ 
are the likelihood measurements from the relevant detectors~\cite{lid}.
We select $\mu^\pm$ candidates with $\mathcal{R}_{\mu} >$ 0.9 (0.97)
if $p_{\mu}>$ 1 GeV/$c$ (0.7$<p_{\mu}<$1.0 GeV/$c$).
These requirements retain about 80\% of muons while removing 98.5\% of pions.
Electron candidates are required to have 
$\mathcal{R}_{e} >$ 0.9, $\mathcal{R}_{\mu} <$ 0.8, and $p_{e}>$ 0.4 GeV/$c$.
These requirements retain about 90\% of electrons while removing 99.7\% of pions.
Bremsstrahlung photons emitted by the electrons are recovered by adding 
neutral clusters found within a 50 mrad cone along the electron direction. 
The energy of the additional photon is required to be less than 0.5 GeV. 

Two oppositely-charged tracks are used to reconstruct $K_{S}^{0} \to \pi^+ \pi^-$ candidates.
The invariant mass is required to be within the range 483--513 MeV/$c^2$ 
($\pm$5 times the $K_{S}^{0}$ reconstructed-mass resolution). 
Other selection criteria are mainly based on 
the distance and the direction of the $K_{S}^{0}$ vertex 
and the distance of daughter tracks to the IP.
For $\pi^{0} \to \gamma\gamma$ candidates, a minimum photon energy of 50 MeV in lab frame is required 
and the invariant mass must be in the range 115 $< M_{\gamma\gamma} < $ 152 MeV/$c^2$ 
($\pm$3 times the $\pi^0$ reconstructed-mass resolution).
Requirements on the photon energy asymmetry, 
$|E_{\gamma}^1 - E_{\gamma}^2|/(E_{\gamma}^1 + E_{\gamma}^2)<0.9$,
and the minimum momentum of the $\pi^0$ candidate in the lab frame, $p_{\pi^0} >$ 200 MeV/$c$,
suppress the combinatorial background. 

$B$-meson candidates are reconstructed by combining a $K^{(*)}$ candidate and a pair of 
oppositely charged leptons, and selected using 
the beam-energy constrained mass $M_{\mathrm{bc}} \equiv \sqrt{E_{\mathrm{beam}}^{2} - p_{B}^{2}}$ 
and the energy difference $\Delta E \equiv E_{B} - E_{\mathrm{beam}}$ 
where $E_{B}$ and $p_{B}$ are the reconstructed energy and momentum
of the $B$ candidate in the $\Upsilon(4S)$ rest frame 
and $E_{\mathrm{beam}}$ is the beam energy in this frame. 
Bremsstrahlung photons are included in the calculation of the momenta of electrons 
and hence are included in the calculations of $M_{\mathrm{bc}}$, $E_{\mathrm{beam}}$ and $q^2$.
We require $B$-meson candidates to be within the region
$M_{\mathrm{bc}}>5.20$~GeV/$c^2$ and $-$35 ($-$55) $<\Delta E< 35$~MeV for the muon (electron) modes.
The signal region is defined by 5.27 $< M_{\mathrm{bc}} <$ 5.29~GeV/$c^2$. 
For $K^*$ modes, the $M_{K\pi}$ candidate (signal) region is defined as 
$M_{K\pi} <$ 1.2~GeV/$c^2$ ($|M_{K\pi}-m_{K^*}|<80$~MeV/$c^2$).

The main backgrounds are continuum $e^+e^- \to q\overline{q}$ ($q=u,d,c,s$) 
and semileptonic $B$ events. 
A Fisher discriminant including 16 modified Fox-Wolfram moments~\cite{sfw} 
is used to exploit the differences between
the event shapes for continuum $q\overline{q}$ production 
(jet-like) and for $B\overline{B}$ decay (spherical) in the $e^+e^-$ rest frame.
We combine 1) the Fisher discriminant,
2) the missing mass $M_{\rm miss} \equiv \sqrt{E_{\rm miss}^{2} - p_{\rm miss}^{2}}$, 
3) the angle between the momentum vector of the reconstructed $B$ candidate and 
the beam direction ($\cos\theta_B$), 
and 4) the distance in the $z$ direction between the candidate $B$ vertex and a vertex position 
formed by the charged tracks that are not associated with the candidate $B$-meson 
into a single likelihood ratio
$\mathcal{R} = {\cal L}_s/({\cal L}_s + {\cal L}_{q\overline{q}})$, where
${\cal L}_s$ (${\cal L}_{q\overline{q}}$) denotes the signal (continuum) likelihood.
For the suppression of semileptonic $B$ decays, 
we combine the Fisher discriminant, $M_{\rm miss}$, $\cos\theta_B$, 
and the lepton separation near the IP in the $z$ direction to 
form the likelihood ratio $\mathcal{R}_{B} = {\cal L}_s/({\cal L}_s + {\cal L}_{B\overline{B}})$, 
where ${\cal L}_{B\overline{B}}$ is the likelihood for semileptonic $B$ decays.

Combinatorial background suppression is improved
by including $q^2$ and $B$-flavor tagging information~\cite{Kakuno:2004cf}, 
which is parameterized by a discrete variable $q_{\rm tag}$ indicating 
the flavor of the tagging $B$-meson candidate and a quality parameter $r$ 
(ranging from 0 for no flavor information to 1 for unambiguous flavor assignment).
Selection criteria for $\mathcal{R}$ and $\mathcal{R}_{B}$ are determined by maximizing
the value of $S/\sqrt{S+B}$, where $S$ and $B$ denote the expected yields 
of signal and background events in the signal region, respectively,
in different ($q^2$,\ $q_{\rm rec} \cdot q_{\rm tag} \cdot r$) regions,
where $q_{\rm rec}$ is the charge of the reconstructed $B$ candidate.
Events with $q_{\rm rec} \cdot q_{\rm tag} \cdot r$ close to $-1$
are considered to be well tagged and are unlikely to be from continuum processes.
For the $K_S^0 \ell^+ \ell^-$ modes, only the dependence on $r$ is considered.

The dominant peaking backgrounds are from the $B \to J/\psi X$ and $\psi^\prime X$ decays 
and are rejected in the $q^2$ regions (the limits are given in units of GeV$^2$/$c^2$): 
\begin{eqnarray}
\nonumber 8.68 < &q^2(\mu^+\mu^-)& < 10.09~, \\
\nonumber 12.86 < &q^2(\mu^+\mu^-)& < 14.18~, \\
\nonumber 8.11 < &q^2(e^+e^-)& < 10.03~, \\
\nonumber 12.15 < &q^2(e^+e^-)& < 14.11~. 
\end{eqnarray}
The decay $B^+ \to J/\psi(\psi^\prime) h^+$ ($h^+ = K^+,\pi^+$) 
can also contribute to the $B^+\to K^+ \pi^-\mu^+\mu^-$ and $K_{S}^{0} \pi^+\mu^+\mu^-$ samples 
if a muon from $J/\psi(\psi^\prime)$ is misidentified as a pion and
another non-muon track is at the same time misidentified as a muon. 
We remove such events from the two samples with the requirement 
$-0.10$ GeV/$c^2$ $< M(\pi\mu)-m_{J/\psi(\psi^\prime)} <$ $0.08$ GeV/$c^2$.
The charmonium $B \to D X$ background can contribute to the muon modes 
if a pion from $D$ meson is misidentified as a muon.
Additional veto windows $|M_{K\mu}-m_{D}|<$ 0.02 GeV/$c^2$ and $|M_{K\pi\mu}-m_{D}|<$ 0.02 GeV/$c^2$ 
suppress this background. 
The invariant mass of the electron pair must exceed 
0.14 GeV/$c^2$ in order to remove 
background from photon conversions and $\pi^0 \to \gamma e^+e^-$ decays. 

If multiple $B$ candidates survive these selections in an event, 
we select the one with the smallest $|\Delta E|$.
The fractions of multiple $B$ events are about 7\%, 12\%, and 20\% for 
the $K^{+} \pi^{-}$, $K_{S}^{0} \pi^{+}$, and $K^{+} \pi^{0}$ modes, 
respectively, according to a Monte Carlo (MC) study.

\begin{figure}[htpb]
\begin{center}
\includegraphics[width=9cm]{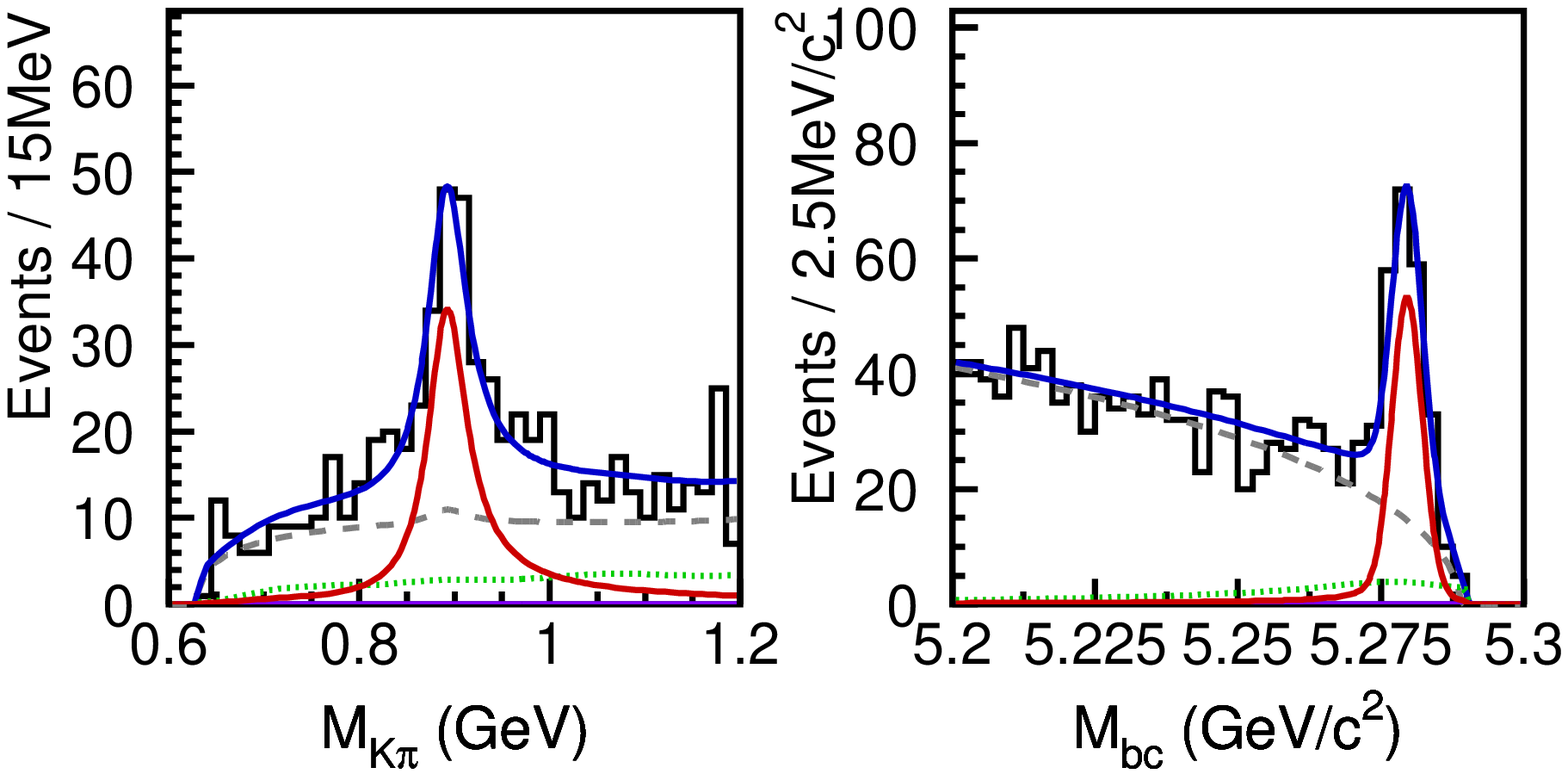} \\
\hskip -0.1cm
\includegraphics[width=4.5cm]{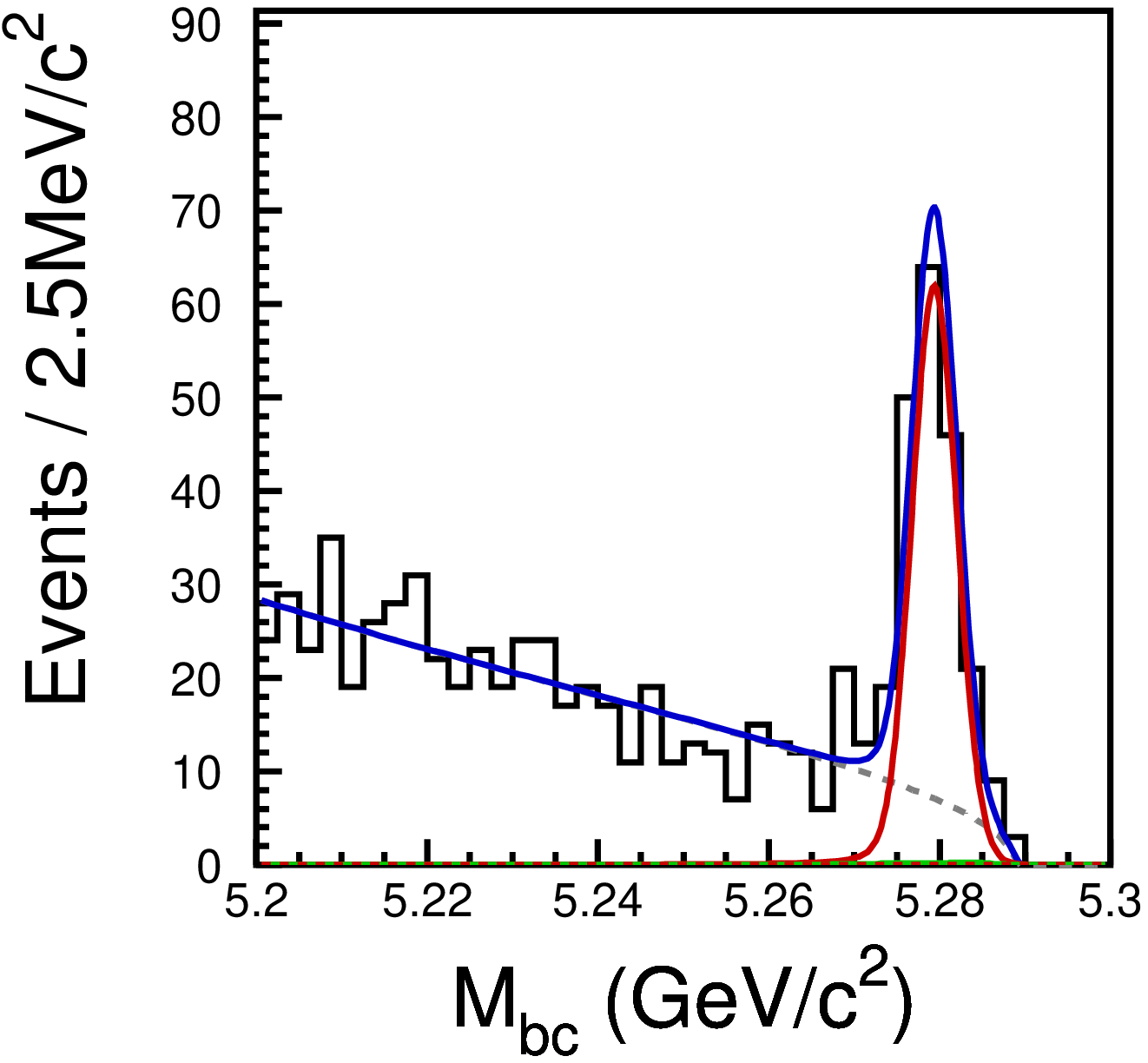}
\hskip -0.3cm
\end{center}
\caption{
Distributions of $M_{K\pi}$ ($M_{\rm bc}$) with fit results superimposed for
the events in the $M_{\rm bc}$ ($M_{K\pi}$) signal region.
The solid curves, solid peak, dashed curves, and dotted curves represent 
the combined fit result, fitted signal, combinatorial background, and $J/\psi(\psi^\prime) X$ background, 
respectively.
}
\label{fig:bfit}
\end{figure}

We perform an extended unbinned maximum likelihood fit 
to $M_{\rm bc}$ and $M_{K\pi}$ in $B \to K^{*} \ell^+ \ell^-$ decays, 
and to $M_{\rm bc}$ in $B \to K \ell^+ \ell^-$ decays. 
The likelihood function is defined as follows:
\begin{eqnarray}
\nonumber
\mathcal{L} & = & {e^{-(N_s+N_b+N_{c\overline{c} X}+N_{K^{(*)}hh})}\over N!} \times \\
\nonumber
&&\prod_{i=1}^{N}~[
N_s P_s^{i}+ N_b P_b^{i} + N_{c\overline{c} X} P_{c\overline{c} X}^{i} +  N_{K^{(*)}hh} P_{K^{(*)}hh}^{i}]~.~~~~~~
\end{eqnarray}
where $N$ denotes the number of observed events in the candidate region, 
and $N_s$ ($P_s^{i}$), $N_b$ ($P_b^{i}$), 
$N_{c\overline{c} X}$ ($P_{c\overline{c} X}^{i}$), 
and $N_{K^{(*)}hh}$ ($P_{K^{(*)}hh}^{i}$) 
denote the event yields (the probability density functions, PDFs, for the i-th event)  
for signal, combinatorial, $B \to J/\psi(\psi^\prime) X$, 
and $B \to K^{(*)}hh$ backgrounds.
The signal PDFs consist of a Gaussian (Crystal Ball function~\cite{cbline}) 
in $M_{\rm bc}$ for the muon (electron) modes and 
a relativistic Breit-Wigner shape in $M_{K \pi}$ for the $K^*$ resonance.
The self-cross-feed PDFs, where the pion or kaon is misidentified, 
are modeled by a two-dimensional smoothed histogram function 
and included in the signal PDFs as well.
The means and widths are determined from MC and calibrated using $B \to J/\psi K^{(*)}$ decays.
The combinatorial PDFs are represented by a product of 
an empirical background function introduced by ARGUS~\cite{argus} in $M_{\rm bc}$, 
and a threshold function, in which the $M_{K \pi}$ threshold is fixed at $m_K + m_{\pi}$, 
plus a relativistic Breit-Wigner shape at the $K^*$ resonance in $M_{K \pi}$.  
The PDFs and yields for $B \to J/\psi(\psi^\prime) X$ decays 
are determined from large MC sample, 
while the $B \to K^{(*)}hh$ PDFs and normalizations are determined from measured data, 
taking into account the probabilities of the pions being misidentified as muons.
Yields for signal and combinatorial background, and the combinatorial PDF parameters
are allowed to float in the fit while the yields and parameters for other components are fixed.
Fig.~\ref{fig:bfit} illustrates the fits for $B$ yields, 
which are 230$^{+24}_{-23}$ and 166$^{+15}_{-15}$ for 
the $K^{*} \ell^+ \ell^-$ and $K \ell^+ \ell^-$ modes, respectively.

\begin{figure}[htb]
\begin{center}
\vskip 0.5cm
\hskip -4.0cm {\bf (a)}\\
\vskip -1.5cm
\includegraphics[width=7cm]{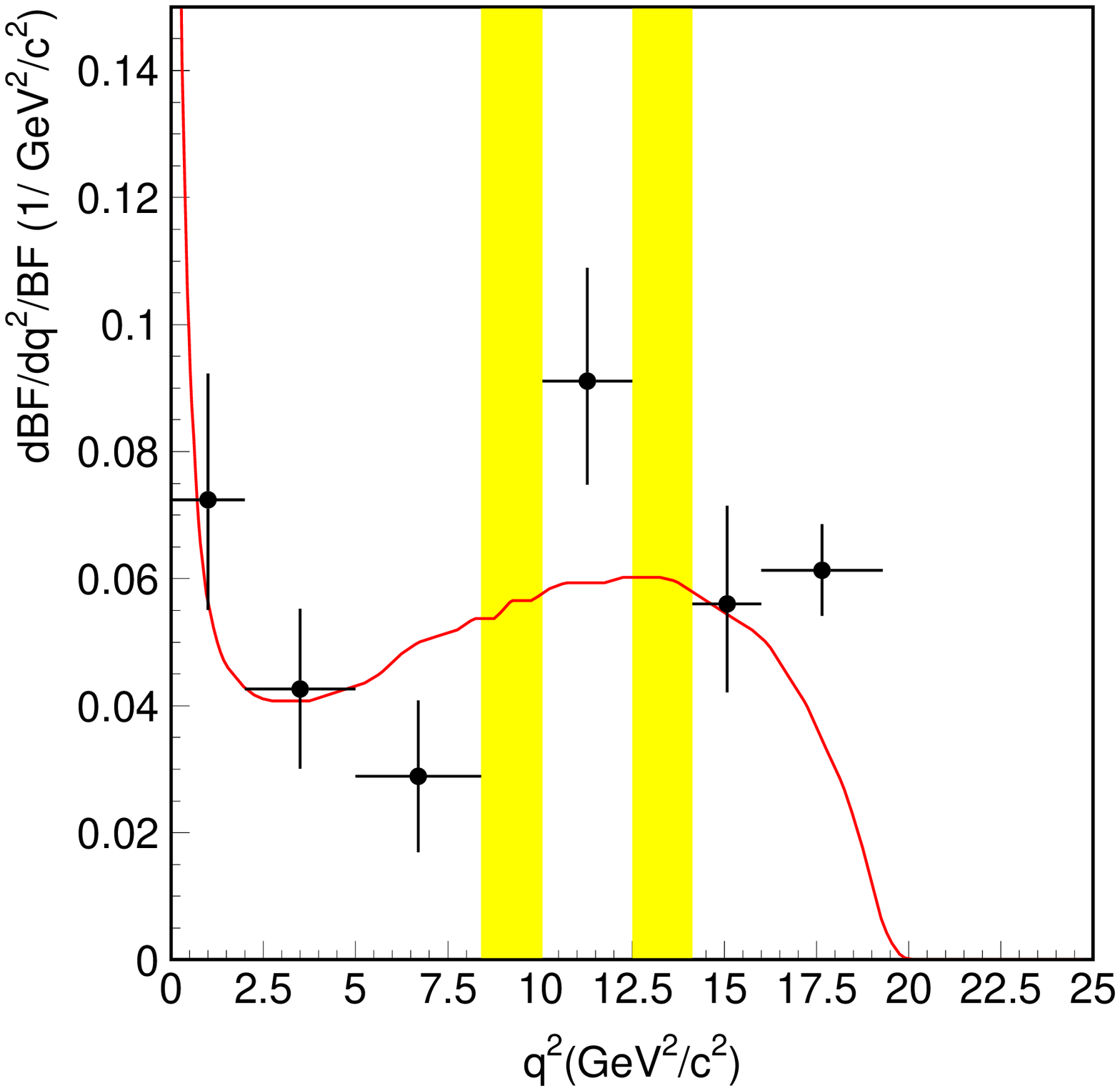}\\
\vskip 0.5cm
\hskip -4cm {\bf (b)}\\
\vskip -1.5cm
\includegraphics[width=7cm]{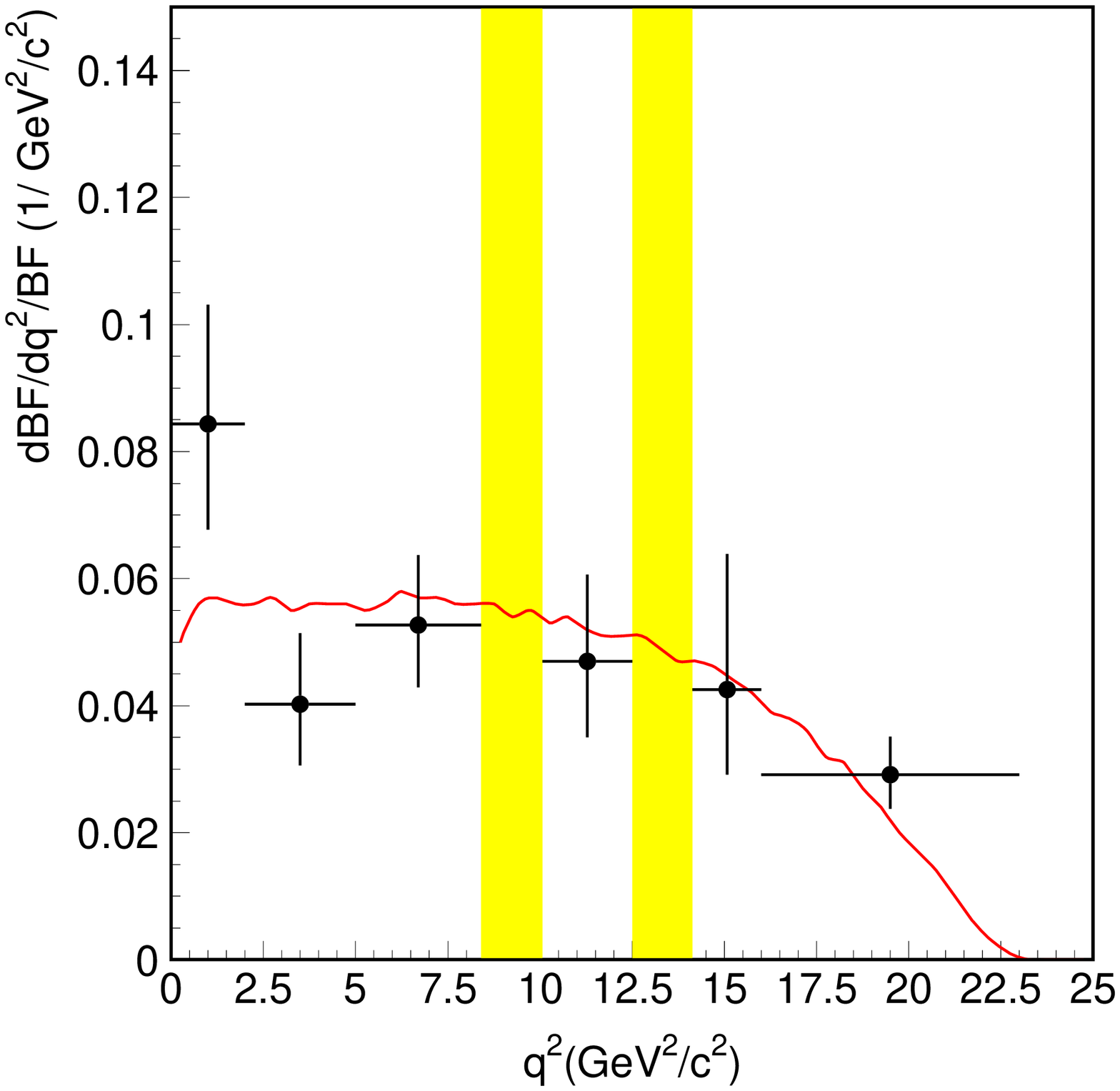}
\end{center}
\vskip -0.5cm
\caption{
Differential branching fractions for 
(a) $K^* \ell^+ \ell^-$ and (b) $K \ell^+ \ell^-$ modes as a function of $q^2$. 
The two shaded regions are veto windows to reject $J/\psi(\psi^\prime) X$ events. 
The solid curve is the theoretical prediction~\cite{Ali:2002jg}.
}
\label{fig:dbf}
\end{figure}

\begin{figure}[t!]
\vskip -0.5cm
\begin{center}
\includegraphics[width=9cm]{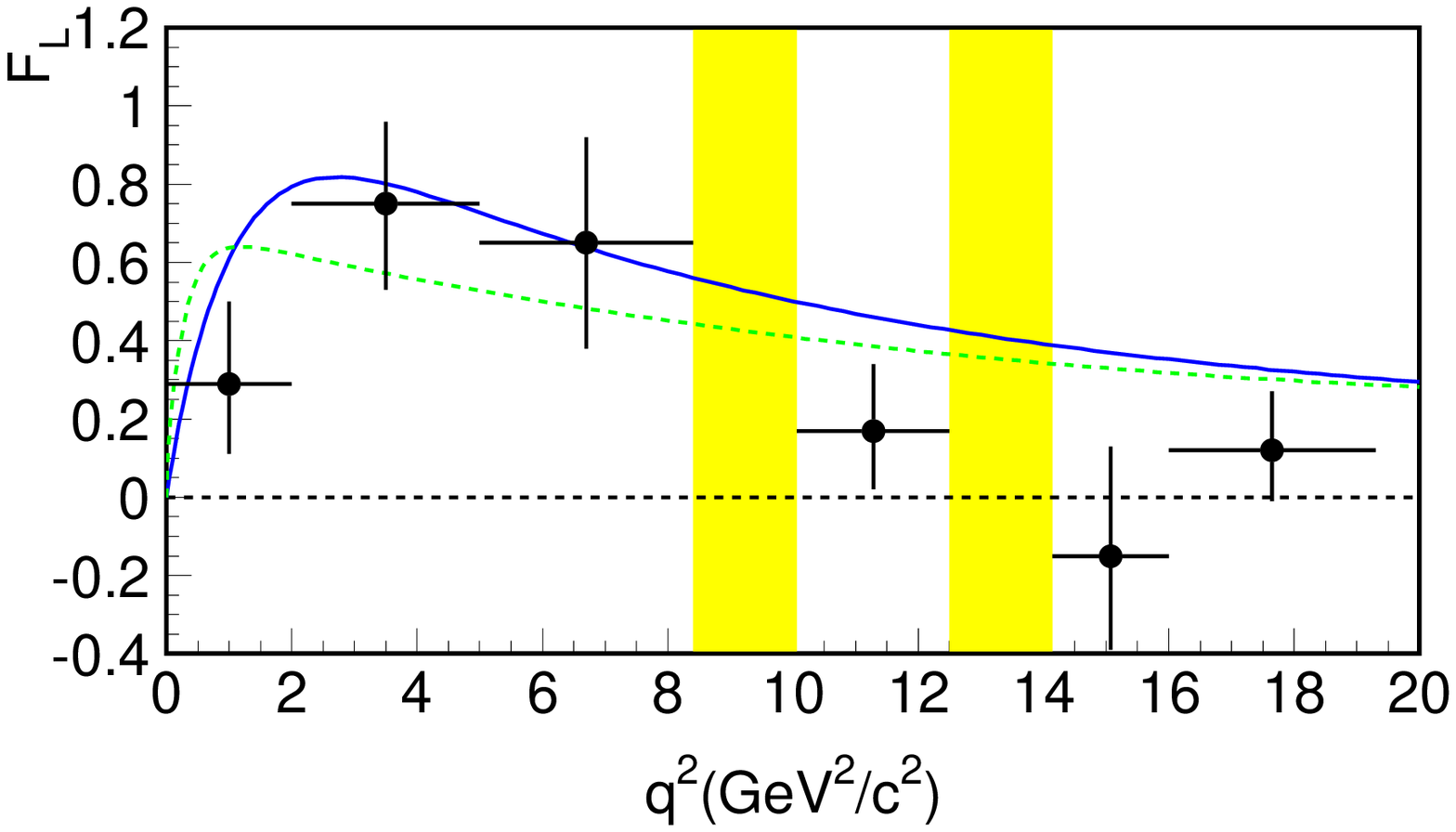}
\end{center}
\vskip -0.5cm
\caption{
Fit results for $F_L$ as a function of $q^2$. 
The solid (dashed) curve shows the SM ($C_7=-C^{SM}_7$) prediction.
}
\label{fig:fl}
\end{figure}

\begin{figure}[t!]
\vskip -0.5cm
\begin{center}
\includegraphics[width=9cm]{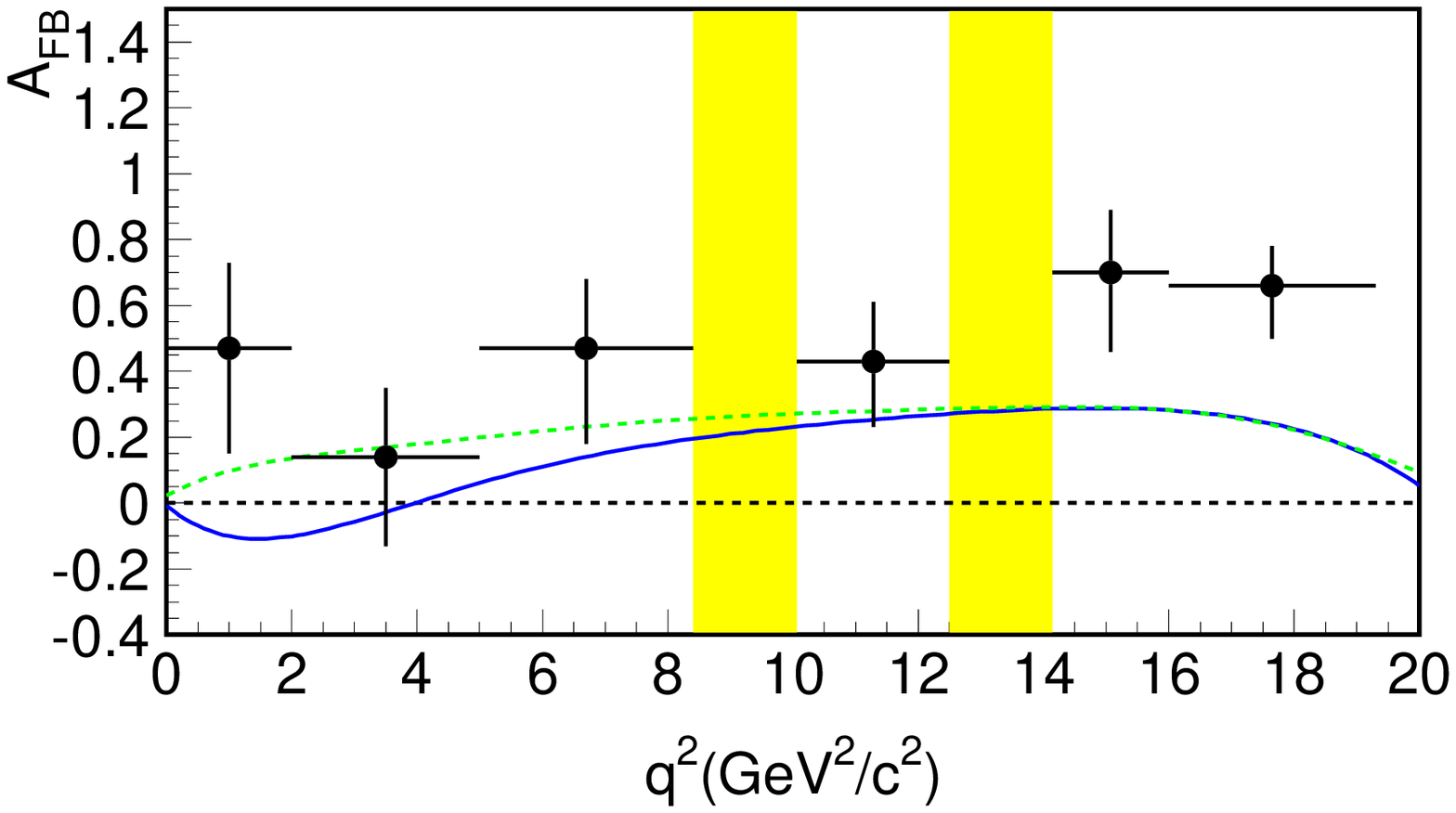}
\end{center}
\vskip -0.5cm
\caption{
Fit results for $A_{FB}$ as a function of $q^2$.
The solid (dashed) curve shows the SM ($C_7=-C^{SM}_7$) prediction.
}
\label{fig:afb}
\end{figure}

We divide $q^2$ into 6 bins and extract the signal and combinatorial background yields in each bin.
The $K^*$ longitudinal polarization fractions ($F_L$) and $A_{FB}$ are 
extracted from fits in the signal region to 
$\cos \theta_{K^*}$ and $\cos \theta_{B\ell}$, respectively, 
where $\theta_{K^*}$ is the angle between the kaon direction and 
the direction opposite the $B$ meson 
in the $K^*$ rest frame, 
and $\theta_{B\ell}$ is the angle between the $\ell^+$ ($\ell^-$) 
and the opposite of the $B$ ($\overline{B}$) direction in the dilepton rest frame. 
The signal PDF for the fit to $\cos \theta_{K^*}$ is described by 
\begin{eqnarray}
\nonumber [~\textstyle{3\over 2} F_L \cos \theta_{K^*} + \textstyle{3\over 4}(1-F_L)(1-\cos \theta_{K^*})~]
\times \epsilon(\cos \theta_{K^*})~,
\end{eqnarray}
where $\epsilon(\cos \theta_{K^*})$ denotes the efficiency obtained from MC. 
For the fit to $\cos \theta_{B\ell}$, we use 
\begin{eqnarray}
\nonumber [~\textstyle{3\over 4} F_L (1-\cos^2 \theta_{B\ell}) + \textstyle{3\over 8}(1-F_L)(1+\cos^2 \theta_{B\ell}) \\
\nonumber + A_{FB}\cos \theta_{B\ell}~] \times \epsilon(\cos \theta_{B\ell})~
\end{eqnarray}
as the signal PDF, 
where $\epsilon(\cos \theta_{B\ell})$ denotes the efficiency 
as a function of $\cos \theta_{B\ell}$. 
The angular efficiency distributions, background PDFs, and signal and background sizes, 
obtained from either MC or a $M_{\rm bc}$--$M_{K\pi}$ fit, 
are fixed in both angular fits. 
$F_L$ ($A_{FB}$) is the only free parameter in the fit to $\cos \theta_{K^*}$ ($\cos \theta_{B\ell}$). Table~\ref{tb:q2_result} lists the measurements of 
$B$ yields, $F_L$, $A_{FB}$, and the partial branching fractions, 
obtained by correcting the $B$ yields for $q^2$ dependent efficiencies.
The differential branching fraction, $F_L$, and $A_{FB}$ as functions of $q^2$ 
for $K^{*} \ell^+ \ell^-$ and $K \ell^+ \ell^-$ modes are shown in 
Fig.~\ref{fig:dbf}, Fig.~\ref{fig:fl}, and Fig.~\ref{fig:afb}, respectively.
The total branching fractions for the entire $q^2$ region, 
obtained by extrapolation from the partial branching fractions, 
as well as the $CP$ asymmetries, 
for the $B \to K^* \ell^+ \ell^-$ and $B \to K \ell^+ \ell^-$ modes 
are listed in Table~\ref{tb:tbf}. 

\begin{table*}[htb]
\caption{
Fit results in each of 6 $q^2$ bins and an additional bin from 1 to 6 GeV$^2$/$c^2$ 
for which recent theory predictions are available~\cite{q2_1-6}.
}
\label{tb:q2_result}
\begin{center}
\begin{tabular}{c|ccccc}
\hline
\hline
$q^2$ (GeV$^2$/$c^2$) & $N_s$ & ${\mathcal B}$(10$^{-7}$) & $A_I$  & $F_L$ & $A_{FB}$ \\
\hline
\multicolumn{6}{c}{$B\to K^* \ell^+ \ell^-$} \\
\hline
0--2         & 27.4$^{+7.4}_{-6.6}$~ & 1.46$^{+0.40}_{-0.35}\pm$0.12~ & $-0.67^{+0.18}_{-0.16}\pm$0.03~ & 
$0.29^{+0.21}_{-0.18}\pm$0.02~  & 0.47$^{+0.26}_{-0.32}\pm$0.03 \\
2--5         & 25.5$^{+7.6}_{-6.8}$~ & 1.29$^{+0.38}_{-0.34}\pm$0.10~ &  $1.17^{+0.72}_{-0.82}\pm$0.02~ & 
$0.75^{+0.21}_{-0.22}\pm$0.05~  & 0.14$^{+0.20}_{-0.26}\pm$0.07 \\
5--8.68      & 20.2$^{+8.3}_{-7.3}$~ & 0.99$^{+0.41}_{-0.36}\pm$0.08~ & $-0.47^{+0.31}_{-0.29}\pm$0.04~ & 
$0.65^{+0.26}_{-0.27}\pm$0.06~  & 0.47$^{+0.16}_{-0.25}\pm$0.14 \\
10.09--12.86 & 54.0$^{+10.5}_{-9.6}$~& 2.24$^{+0.44}_{-0.40}\pm$0.18~ &  $0.00^{+0.20}_{-0.21}\pm$0.05~ & 
$0.17^{+0.17}_{-0.15}\pm$0.03~  & 0.43$^{+0.18}_{-0.20}\pm$0.03 \\
14.18--16    & 36.2$^{+9.9}_{-8.8}$~ & 1.05$^{+0.29}_{-0.26}\pm$0.08~ &  $0.16^{+0.30}_{-0.35}\pm$0.05~ & 
$-0.15^{+0.27}_{-0.23}\pm$0.07~ & 0.70$^{+0.16}_{-0.22}\pm$0.10 \\
$>$16        & 84.4$^{+11.0}_{-9.9}$~& 2.04$^{+0.27}_{-0.24}\pm$0.16~ & $-0.02^{+0.20}_{-0.21}\pm$0.05~ & 
0.12$^{+0.15}_{-0.13}\pm$0.02~  & 0.66$^{+0.11}_{-0.16}\pm$0.04 \\
\hline
1--6         & 29.42$^{+8.9}_{-8.0}$~ & 1.49$^{+0.45}_{-0.40}\pm$0.12~ &  $0.33^{+0.37}_{-0.43}\pm$0.05~ & 
$0.67^{+0.23}_{-0.23}\pm$0.05~  & 0.26$^{+0.27}_{-0.30}\pm$0.07 \\
\hline
\multicolumn{6}{c}{$B\to K \ell^+ \ell^-$} \\
\hline
0--2         & 27.0$^{+6.0}_{-5.4}$~ & 0.81$^{+0.18}_{-0.16}\pm$0.05~ & $-0.33^{+0.33}_{-0.25}\pm$0.05 
&$-$& $0.06^{+0.32}_{-0.35}\pm$0.02  \\
2--5         & 22.5$^{+6.0}_{-5.3}$~ & 0.58$^{+0.16}_{-0.14}\pm$0.04~ & $-0.49^{+0.45}_{-0.34}\pm$0.04 
&$-$& $-0.51^{+0.31}_{-0.31}\pm$0.09  \\
5--8.68      & 34.1$^{+7.1}_{-6.5}$~ & 0.86$^{+0.18}_{-0.16}\pm$0.05~ & $-0.19^{+0.26}_{-0.22}\pm$0.05 
&$-$& $-0.18^{+0.12}_{-0.15}\pm$0.03  \\
10.09--12.86 & 22.0$^{+6.2}_{-5.5}$~ & 0.55$^{+0.16}_{-0.14}\pm$0.03~ & $-0.29^{+0.37}_{-0.29}\pm$0.05 
&$-$& $-0.21^{+0.17}_{-0.15}\pm$0.06  \\
14.18--16    & 15.6$^{+4.9}_{-4.3}$~ & 0.38$^{+0.19}_{-0.12}\pm$0.02~ & $-0.40^{+0.61}_{-0.69}\pm$0.04 
&$-$& $0.04^{+0.32}_{-0.26}\pm$0.05  \\
$>$16        & 40.3$^{+8.2}_{-7.5}$~ & 0.98$^{+0.20}_{-0.18}\pm$0.06~ &  $0.11^{+0.24}_{-0.21}\pm$0.05 
&$-$& $0.02^{+0.11}_{-0.08}\pm$0.02  \\
\hline
1--6         & 52.0$^{+8.7}_{-8.0}$~ & 1.36$^{+0.23}_{-0.21}\pm$0.08~ & $-0.41^{+0.25}_{-0.20}\pm$0.04  
&$-$& $-0.04^{+0.13}_{-0.16}\pm$0.05 \\
\hline
\hline
\end{tabular}
\end{center}
\end{table*} 

\begin{table}[htb]
\caption{
Total branching fractions for $B\to K^* \ell^+ \ell^-$ and $B\to K\ell^+\ell^-$ decays. 
}
\label{tb:tbf}
\begin{center}
\begin{tabular}{c|cc}
\hline
\hline
Mode & ${\mathcal B}$ (10$^{-7}$)  & $A_{CP}$  \\
\hline
$K^{*+}\mu\mu$ & $11.4^{+3.2}_{-2.7}$$\pm$1.0~ & $-0.12^{+0.24}_{-0.24}$$\pm$0.02 \\
$K^{*0}\mu\mu$ & $10.8^{+1.9}_{-1.5}$$\pm$0.7~ & $0.00^{+0.15}_{-0.15}$$\pm$0.03 \\
$K^{*}\mu\mu$  & $11.2^{+1.6}_{-1.4}$$\pm$0.8~ & $-0.03^{+0.13}_{-0.13}$$\pm$0.02 \\
\hline
$K^{*+}ee $ & $16.4^{+5.0}_{-4.2}$$\pm$1.8~ & $-0.14^{+0.23}_{-0.22}$$\pm$0.02 \\
$K^{*0}ee $ & $11.8^{+2.6}_{-2.1}$$\pm$0.9~ & $-0.21^{+0.19}_{-0.19}$$\pm$0.02 \\
$K^{*}ee $  & $13.7^{+2.3}_{-2.0}$$\pm$1.2~ & $-0.18^{+0.15}_{-0.15}$$\pm$0.01 \\
\hline
$K^{*+}\ell\ell$ & $12.4^{+2.3}_{-2.0}$$\pm$1.2~ & $-0.13^{+0.17}_{-0.16}$$\pm$0.01 \\
$K^{*0}\ell\ell$ &  $9.8^{+1.3}_{-1.1}$$\pm$0.7~ & $-0.08^{+0.12}_{-0.12}$$\pm$0.02 \\
$K^{*}\ell\ell$  & $10.8^{+1.1}_{-1.0}$$\pm$0.9~ & $-0.10^{+0.10}_{-0.10}$$\pm$0.01 \\
\hline
\hline
$K^{+}\mu\mu$ & $5.3^{+0.8}_{-0.7}$$\pm$0.3~ & $-0.05^{+0.13}_{-0.13}$$\pm$0.03 \\
$K^{0}\mu\mu$ & $4.3^{+1.3}_{-1.0}$$\pm$0.2~ & $-$ \\
$K\mu\mu$     & $5.0^{+0.6}_{-0.6}$$\pm$0.3~ & $-$ \\
\hline
$K^{+}ee $ & $5.7^{+0.9}_{-0.8}$$\pm$0.3~ & $-0.14^{+0.14}_{-0.14}$$\pm$0.03 \\
$K^{0}ee $ & $2.0^{+1.4}_{-1.0}$$\pm$0.1~ &  $-$ \\
$Kee $     & $4.8^{+0.8}_{-0.7}$$\pm$0.3~ &  $-$ \\
\hline
$K^{+}\ell\ell$ & $5.3^{+0.6}_{-0.5}$$\pm$0.3~ & $-0.04^{+0.10}_{-0.10}$$\pm$0.02 \\
$K^{0}\ell\ell$ & $3.3^{+0.9}_{-0.7}$$\pm$0.2~ & $-$ \\
$K\ell\ell$     & $4.8^{+0.5}_{-0.4}$$\pm$0.3~ & $-$ \\
\hline
\hline
\end{tabular}
\end{center}
\end{table} 

\begin{figure}[htb]
\vskip -0.5cm
\begin{center}
\includegraphics[width=9cm]{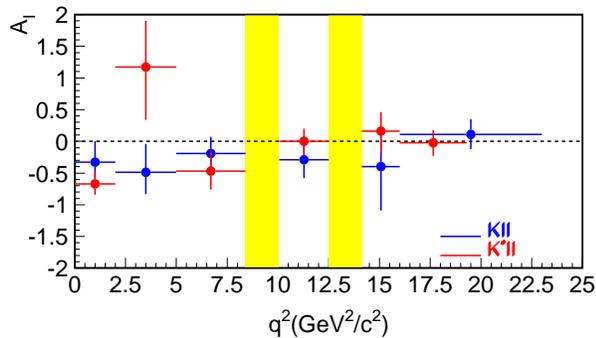}
\end{center}
\vskip -0.5cm
\caption{
$A_I$ as a function of $q^2$ for $K^* \ell^+ \ell^-$ (red) 
and $K \ell^+ \ell^-$ (blue) modes.}
\label{fig:ai}
\end{figure}

\begin{table*}[t!]
\caption{
Systematic uncertainties in the branching fraction (in \%) for each decay channel 
of electron/muon mode. 
}
\label{tb:systematics}
\begin{center}
\begin{tabular}{l|ccc|cc}
\hline\hline
Source & $(K^+\pi^-)^{*0}$ & $(K_S^0\pi^+)^{*+}$ & $(K^+\pi^0)^{*+}$ & $K^+$ & $K_S$ \\
\hline
Tracking           & 4.4 / 4.4 & 3.3 / 3.3 & 3.3 / 3.3 & 3.0 / 3.0 & 2.0 / 2.0 \\
$K$ or $K_S^0$     & 1.0 / 1.0 & 4.9 / 4.9 & 1.0 / 1.0 & 1.0 / 1.0 & 4.9 / 4.9 \\
$\pi$ or $\pi^0$   & 1.0 / 1.0 & 1.0 / 1.0 & 4.0 / 4.0 & $-$ / $-$ & $-$ / $-$ \\
e/$\mu$            & 2.6 / 3.0 & 2.6 / 3.0 & 2.6 / 3.0 & 2.6 / 3.0 & 2.6 / 3.0 \\
$\mathcal{R}$ and 
$\mathcal{R}_{B}$  & 2.1 / 2.0 & 1.9 / 3.6 & 3.0 / 3.4 & 2.0 / 2.5 & 2.5 / 1.2 \\
MC model           & 2.5 / 2.9 & 2.1 / 2.9 & 4.6 / 2.1 & 0.9 / 1.0 & 2.9 / 2.6 \\
Fitting PDF        & 1.6 / 2.1 & 3.0 / 3.3 & 4.3 / 4.6 & 1.4 / 2.0 & 2.5 / 2.1 \\
$B\bar{B}$ pairs   & 1.4 / 1.4 & 1.4 / 1.4 & 1.4 / 1.4 & 1.4 / 1.4 & 1.4 / 1.4 \\
Rare $B$           & 0.6 / 0.5 & 0.9 / 0.9 & 0.3 / 0.3 & 1.3 / 1.0 & 1.3 / 0.8 \\
$c\overline{c}X$   & 1.1 / 1.4 & 1.8 / 2.0 & 4.2 / 7.8 & 0.3 / 0.4 & 0.1 / 0.3 \\
\hline
Total              & 6.5 / 7.3 & 8.0 / 10.1& 9.8 / 12.0& 5.1 / 5.8 & 7.4 / 7.2 \\
\hline\hline
\end{tabular}
\end{center}
\end{table*}

We calculate the ratios of branching fractions for the electron mode to the muon mode.
The lepton flavor ratio for $B\to K^* \ell^+ \ell^-$ ($R_{K^*}$) is 
sensitive to the size of the photon pole and is predicted to be 1.33 in the SM, 
while the ratio for $B\to K \ell^+ \ell^-$ ($R_K$) is sensitive to Higgs emission 
and predicted to be larger than 1.0 in the Higgs doublet model with large tan$\beta$~\cite{susy}.
The results are 
\begin{eqnarray}
\nonumber
R_{K^*} & = & 1.21\pm0.25\pm0.08~,\\
\nonumber
R_{K} & = & 0.97\pm0.18\pm0.06~.
\end{eqnarray}
Assuming the ratios of branching fractions for the electron mode to the muon mode is 
1.33 (1.0) in the $K^*$($K$) mode, 
the combined branching fractions are measured to be 
\begin{eqnarray}
\nonumber
\mathcal{B}(B \to K^* \ell^+ \ell^-) & = & (10.8^{+1.1}_{-1.0}\pm0.9) \times 10^{-7}~,\\
\nonumber
\mathcal{B}(B \to K \ell^+ \ell^-) & = & (4.8^{+0.5}_{-0.4}\pm0.3) \times 10^{-7}~. 
\end{eqnarray}
Isospin asymmetry, shown in Table~\ref{tb:q2_result} and Fig.~\ref{fig:ai}, 
is defined as 
\begin{eqnarray}
\nonumber
A_I \equiv \frac{(\tau_{B^+}/\tau_{B^0}) \times {\mathcal B}(K^{(*)0} \ell^+ \ell^-) - {\mathcal B}(K^{(*)\pm} \ell^+ \ell^-)}
{(\tau_{B^+}/\tau_{B^0}) \times {\mathcal B}(K^{(*)0} \ell^+ \ell^-) + {\mathcal B}(K^{(*)\pm} \ell^+ \ell^-)}~,
\end{eqnarray}
where $\tau_{B^+}/\tau_{B^0}$ (=1.071) is the lifetime ratio of $B^+$ to $B^0$~\cite{pdg}.
A large isospin asymmetry for $q^2$ below the $J/\psi$ resonance was reported recently~\cite{ai_exp}. 
We also measure the combined $A_I$ for $q^2<$ 8.68 GeV$^2$/$c^2$ and find  
\begin{eqnarray}
\nonumber
A_I(B \to K^* \ell^+ \ell^-) & = -0.29^{+0.16}_{-0.16} \pm0.03~& ~\sigma = 1.40~,\\
\nonumber
A_I(B \to K \ell^+ \ell^-) & = -0.31^{+0.17}_{-0.14} \pm0.05~& ~\sigma = 1.75~,\\
\nonumber
A_I(B \to K^{(*)} \ell^+ \ell^-) & = -0.30^{+0.12}_{-0.11} \pm0.04~& ~\sigma = 2.24~,
\end{eqnarray}
where $\sigma$ denotes the significance from null asymmetry 
and is defined as $\sigma \equiv \sqrt{-\rm{2ln} \left( \mathcal{L}_{0} / \mathcal{L}_{\rm max} \right)}$, 
where $\mathcal{L}_{0}$ is the likelihood with $A_I$ constrained to be zero 
and $\mathcal{L}_{\rm max}$ is the maximum likelihood. 
Systematic uncertainties are considered in the significance calculation.

\begin{table}[htb]
\caption{
Systematic errors on $F_L$ and $A_{FB}$ measurements.}
\label{tb:angsys}
\begin{center}
\begin{tabular}{l|cc}
\hline\hline
Source & $F_L$ & $A_{FB}$ \\
\hline
Signal yield  & 0.01--0.06 & 0.00--0.06 \\ 
Background    & 0.01--0.03 & 0.01--0.03 \\
$F_L$         &    $-$     & 0.01--0.13 \\
Fitting bias  &    0.01    &    0.02    \\
Fitting PDF   &    0.01    &    0.01    \\
\hline\hline
\end{tabular}
\end{center}
\end{table}

Systematic uncertainties in the branching fraction measurement 
for each decay channel are summarized in Table~\ref{tb:systematics}.
They stem dominantly from 
tracking efficiencies (2.0\%--4.4\%), MC decay models (0.9\%--4.6\%), 
electron (3.0\%) and muon (2.6\%) identification, 
$K^0_S$ (4.9\%) and $\pi^0$ (4.0\%) reconstruction, 
and $\mathcal{R}$ and $\mathcal{R}_{B}$ selection (1.2\%--3.6\%).
The signal MC samples are generated based on a decay model derived from~\cite{Ali:2002jg}, 
and the modeling uncertainties are evaluated by comparing different MC samples 
based on different decay models~\cite{model}, 
while lepton identification is studied using a $J/\psi \to \ell^+ \ell^-$ data control sample. 
For $\mathcal{R}$ and $\mathcal{R}_{B}$ selections, 
we estimate the uncertainties from large control samples with the same final states, 
$B \to J/\psi K^{(*)}$ with $J/\psi \to \ell^+ \ell^-$.
Other uncertainties such as kaon and pion identification efficiencies, 
fitting PDFs, background contamination from $J/\psi$ decays and charmless $B$ decays, 
and the number of $B\overline{B}$ pairs are found to be small.
The systematic error on $R_{K^{(*)}}$ ($A_I$) is determined by combining 
the uncertainties from lepton ($K$/$\pi$) identification, 
$\mathcal{R}$ and $\mathcal{R}_{B}$ selections, fitting PDFs 
and background contamination. 
The uncertainty in $A_I$ from the assumption 
of equal production of $B^0$ and $B^+$ is also considered.
Table~\ref{tb:angsys} shows the systematic uncertainties for angular fits. 
The main uncertainties are propagated from the errors on the fixed normalizations and $F_L$, 
determined from $M_{\rm bc}$--$M_{K\pi}$ and $\cos \theta_{K^*}$ fits, respectively. 
Fitting bias and fitting PDFs are checked using large $B \to J/\psi K^{(*)}$ and MC samples. 
The total uncertainties range from 0.02--0.06 and 0.03--0.15 
for $F_L$ and $A_{FB}$ fits, respectively. 
The systematic errors on $A_{CP}$ 
are assigned using the measured $CP$ asymmetry for sideband data 
without $\mathcal{R}$ and $\mathcal{R}_{B}$ selections and are found to be around 0.01--0.03.

In summary, 
we report the differential branching fraction and isospin asymmetry as functions of $q^2$, 
lepton flavor ratios, and $CP$ asymmetries 
in both $B\to K^* \ell^+ \ell^-$ and $B\to K \ell^+ \ell^-$ decays.
$K^*$ longitudinal polarization and forward-backward asymmetry 
as functions of $q^2$ in $B\to K^* \ell^+ \ell^-$ 
are also measured from an angular analysis. 
The differential branching fraction, lepton flavor ratios, and $K^*$ polarization 
are in good agreement with the SM predictions.
No significant $CP$ asymmetry or isospin asymmetry is found.
The $A_{FB}(q^2)$ spectrum, although consistent with previous measurements~\cite{kll_exp}, 
tends to be shifted toward the positive side from the SM expectation, 
especially at large $q^2$ values.
A much larger data set, 
such as will be available from the proposed super $B$ factory~\cite{superB} and LHCb~\cite{LHCb}, 
is needed to 
make more precise comparisons with the SM and other theoretical predictions.

We thank the KEKB group for the excellent operation of the
accelerator, the KEK cryogenics group for the efficient
operation of the solenoid, and the KEK computer group and
the National Institute of Informatics for valuable computing
and SINET3 network support. We acknowledge support from
the Ministry of Education, Culture, Sports, Science, and
Technology of Japan and the Japan Society for the Promotion
of Science; the Australian Research Council and the
Australian Department of Education, Science and Training;
the National Natural Science Foundation of China under
Contracts No.~10575109 and 10775142; the Department of
Science and Technology of India; 
the BK21 program of the Ministry of Education of Korea, 
the CHEP SRC program and Basic Research program 
(Grant No.~R01-2005-000-10089-0) of the Korea Science and
Engineering Foundation, and the Pure Basic Research Group 
program of the Korea Research Foundation; 
the Polish State Committee for Scientific Research; 
the Ministry of Education and Science of the Russian
Federation and the Russian Federal Agency for Atomic Energy;
the Slovenian Research Agency;  the Swiss
National Science Foundation; the National Science Council
and the Ministry of Education of Taiwan; and the U.S.\
Department of Energy.

\end{document}